\documentclass[aps,pre,showpacs,twocolumn]{revtex4}
\usepackage{graphicx}

\begin{document}

\title{Stabilization of unstable steady states by variable delay feedback control}

\author{Aleksandar Gjurchinovski}

\email{agjurcin@iunona.pmf.ukim.edu.mk}

\author{Viktor Urumov}

\email{urumov@iunona.pmf.ukim.edu.mk}

\affiliation{Department of Physics, Faculty of Natural Sciences
and Mathematics, Sts.\ Cyril and Methodius University,
P.\ O.\ Box 162, 1000 Skopje, Macedonia}

\begin{abstract}
We report on a dramatic improvement of the performance of the
classical time-delayed autosynchronization method (TDAS) to
control unstable steady states, by applying a time-varying delay
in the TDAS control scheme in a form of a deterministic or
stochastic delay-modulation in a fixed interval around a nominal
value $T_0$. The successfulness of this variable delay feedback
control (VDFC) is illustrated by a numerical control simulation of
the Lorenz and R\"{o}ssler systems using three different types of
time-delay modulations: a sawtooth wave, a sine wave, and a
uniform random distribution. We perform a comparative analysis
between the VDFC method and the standard TDAS method for a
sawtooth-wave modulation by analytically determining the domains
of control for the generic case of an unstable fixed point of
focus type.

\end{abstract}


\date{May 22, 2008}

\maketitle

\section{Introduction}

The idea of controlling chaos has been initiated by Ott, Grebogy
and Yorke in a seminal paper proposing a routine to stabilize
unstable orbits embedded in chaotic attractors \cite{OGY90}. This
control scheme, now recognized as the OGY method, is taking
advantage of the ergodicity of the system by applying a small
perturbation to a suitably chosen control parameter when the
trajectory of the system is sufficiently close to the stable
manifold of the local linear system. Since the OGY paper, the
control of chaos has become a topic of broad interest among
scientists, both theoretically and experimentally
\cite{SCH99,BGL00,KAP96,ABC98}. Although different control schemes
were developed, most of them are relying on the general concept
underlying OGY scheme -- stabilizing an orbit embedded into the
chaotic system.

An alternative and very practical implementation of the OGY idea
is the time-delayed autosynchronization (TDAS) introduced by
Pyragas in 1992 \cite{PYR92,PT93,PYR06}. It is based on a
continuous feedback applied in the form of a control force
proportional to the difference of the current state of the system
at time $t$ and its counterpart at some instant $t-T$ in the past.
If the delay time $T$ is set equal to the period of the unstable
periodic orbit (UPO) whose stabilization is required, the control
signal vanishes when the system is on that orbit, so that the
extracted orbit remains a solution of the equations describing the
dynamics of the original system. In this sense, the Pyragas method
is non-invasive.

A generalization of Pyragas' method was suggested by Socolar,
Sukow and Gauthier in 1994 \cite{SSG94,PYR95}, where the feedback
signal was taken in the form of a geometric sum (extended
time-delayed autosynchronization -- ETDAS), or a mean value of a
finite number of delay terms (N time-delayed autosynchronization
-- NTDAS), each using information from many previous states of the
system involving integer multiples of the delay time $T$. In spite
of introducing an additional control parameter, ETDAS has been
recognized as the most important modification of the TDAS, since
it achieves stabilization of UPOs with a higher degree of
instability.

Parallel to the efforts of stabilizing UPOs, the Pyragas' method
and its various extensions were used to stabilize unstable steady
states (USS) \cite{HS05,YWH06,DHS07,PPK02,PPK04}. Among these is
the multiple delay feedback control (MDFC) recently suggested by
Albhorn and Parlitz \cite{AP04,AP05}. The scheme is a natural
extension of ETDAS method, using two or more delayed feedback
signals with incommensurate delay times. Although being more
effective in comparison to TDAS and ETDAS, the MDFC has some
drawback by introducing additional control parameters with every
additional feedback signal.

To improve the performance of TDAS in controlling unstable steady
states, with this paper we suggest a variable delay feedback
control (VDFC), where the delay time $T$ is not kept constant
during the control process, but it is modulated in time in a
suitably chosen way. We demonstrate the VDFC method for the
paradigmatic Lorenz and R\"{o}ssler systems, using three different
types of modulation of the delay time $T(t)$ around a nominal
value $T_0$ in a fixed interval
$[T_0-\varepsilon,T_0+\varepsilon]$ determined by the control
parameter $\varepsilon$. To make a comparison with the TDAS
method, we perform a linear stability analysis of the VDFC for
stabilizing a two-dimensional unstable steady state of a focus
type for different values of the parameter $\varepsilon$. 
By using the same method of analysis, we calculated the parameter
regions for successful VDFC stabilization of the unstable steady states 
in the three-dimensional Lorenz and R\"{o}ssler systems.
We show that for $\varepsilon=0$ VDFC is reduced to TDAS, and 
for $\varepsilon>0$ the domain of control is drastically enlarged 
with respect to TDAS. The correctness of the stability analysis is 
supported by computer simulations.

\section{Variable delay feedback control}

The Pyragas' method assumes a dynamical system having at least one
scalar variable $y(t)$ accessible for measurements, and an input
channel through which the external force $F(t)$ is fed back to the
system. The standard TDAS control scheme reads:
\begin{eqnarray}
{d\over dt}\mathbf{x}(t)&=&\mathbf{Q}(y(t),\mathbf{x}(t)),\\
{d\over dt}y(t)&=&P(y(t),\mathbf{x}(t))+F(t),
\end{eqnarray}
where
\begin{equation}
F(t)=K[y(t-T)-y(t)]
\label{eq:1.3}
\end{equation}
is the external feedback force which is reinjected into the system
through the $y$-channel as the difference between the signal
$y(t)$ and the delayed signal $y(t-T)$ multiplied by a constant
weighting factor $K$. The remaining variables of the system are
represented by the vector $\mathbf{x}$, and $P$ and $\mathbf{Q}$
are nonlinear functions. The scheme involves only two control
parameters -- the delay time $T$ and the control gain $K$. For an
effective stabilization of USSs, the delay time $T$ is related to
the intrinsic characteristic timescale given by the imaginary part
of the stationary point eigenvalue \cite{HS05}.

The external feedback force (\ref{eq:1.3}) in the case of VDFC reads:
\begin{equation}
F(t)=K[y(t-T(t))-y(t)],
\label{eq:1.4}
\end{equation}
where $T(t)$ is the time-dependent delay time. The schematic of VDFC
is shown in Fig. 1.

\begin{figure}
\includegraphics[width=0.9\columnwidth,height=!]{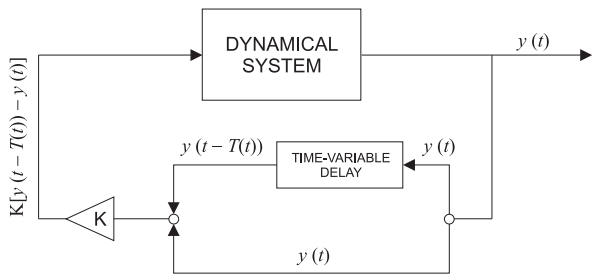}
\caption{Schematic of the variable delay feedback control}
\end{figure}

To illustrate the VDFC method for stabilizing unstable steady states,
we will use the Lorenz system \cite{LOR63}:
\begin{eqnarray}
{d\over dt}x(t)&=&-10\, x(t)+10\, y(t), \label{eq:lor1}\\
{d\over dt}y(t)&=&28\, x(t)-y(t)-x(t)z(t)+F(t), \label{eq:lor2}\\
{d\over dt}z(t)&=&-8/3\, z(t)+x(t)y(t) \label{eq:lor3},
\end{eqnarray}
assuming that the output variable is $y(t)$ and that the external
feedback signal (\ref{eq:1.4}) perturbs only the second equation
of the system. The trajectory of the unperturbed system ($K=0$) is
shown in panel (a) of Fig. 2, where the appearance of the chaotic
attractor becomes evident.

\begin{figure*}
\includegraphics[width=\textwidth,height=!]{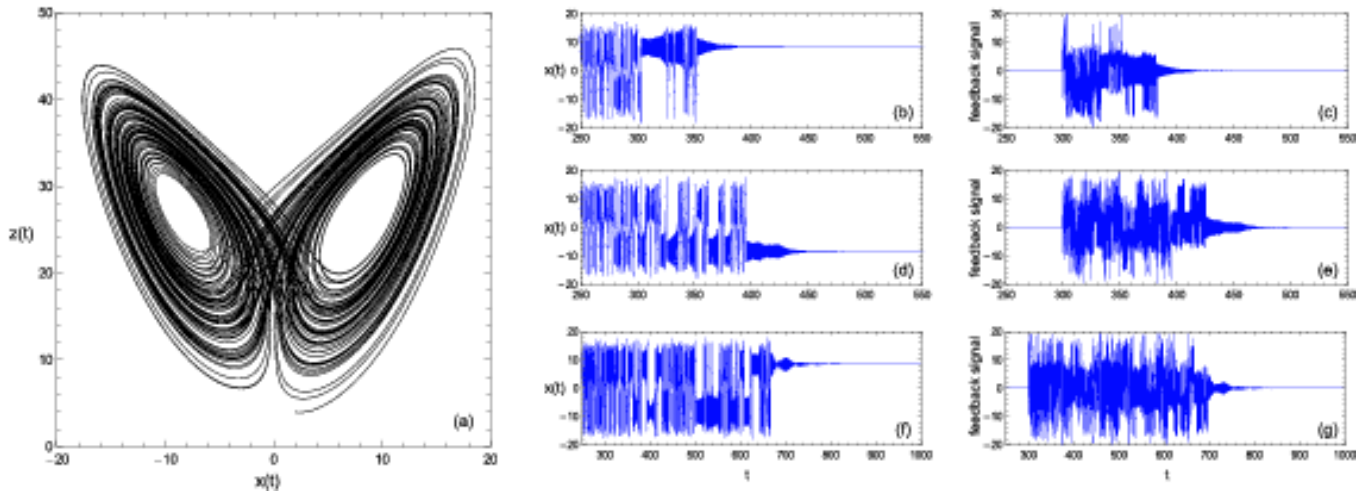} \caption{
(Color online) A simulation of variable delay feedback control in
the Lorenz system (\ref{eq:lor1})--(\ref{eq:lor3}) 
using three different modulations of the delay
time. (a) The Lorenz attractor of the unperturbed system; (b),(c)
The time-series of the variable $x(t)$ and the feedback signal
$F(t)$ of the controlled system corresponding to a sawtooth-wave
modulation with $a=10$, indicating a successful control of the
unstable fixed point at $C^+$; (d),(e) The corresponding time-series 
for a uniform random modulation controlling the unstable point 
at $C^-$; (f),(g) The time-series for a sine-wave modulation
($a=10$), stabilizing the unstable point at $C^+$. In each case,
the initial conditions were: $x(0)=2$, $y(0)=3$ and $z(0)=4$, the
control parameters were: $K=0.5$, $T_0=30$ and $\varepsilon=1$,
and the control was activated at $t=300$. Note the different
scales on the $t$-axis on different panels.}
\end{figure*}

The Lorenz system has three equilibrium points, one at the origin
$C_0(0,0,0)$, and two positioned symmetrically with respect to the
$z$-axis at $C^+(6\sqrt{2},6\sqrt{2},27)$ and
$C^-(-6\sqrt{2},-6\sqrt{2},27)$. Linearizing the Lorenz equations
around the equilibrium points reveals the form of their stability.
The eigenvalues of the unperturbed fixed point $C_0$ at the origin
are $\lambda(C_0)\approx\{-22.83,11.83,-2.67\}$ indicating an
unstable three-dimensional saddle with one positive and two
negative real eigenvalues. The other two fixed points have the
same type of stability given by their common set of eigenvalues
$\lambda(C^{\pm})\approx\{-13.85,0.09+10.19i,0.09-10.19i\}$. They
are unstable fixed points of the focus type, having one negative 
real eigenvalue and complex-conjugate pair of eigenvalues with 
positive real part.

Panels (b)--(g) of Fig. 2 depict the dynamics of the variable
$x(t)$ and the feedback signal $F(t)$ obtained from a computer
simulation of the VDFC-controlled Lorenz system (5)--(7),
indicating a successful control of the fixed points $C^{\pm}$ by
using three different types of modulation of the delay time $T(t)$
in an $\varepsilon$-neighbourhood around a fixed value $T_0$. The
initial conditions of the system were chosen $x(0)=2$, $y(0)=3$
and $z(0)=4$. The control was activated at $t=300$. Panels
(b)--(c) correspond to time-modulation in a form of a sawtooth
wave with an inverse period $a$:
\begin{equation}
T(t)=T_0+\varepsilon\,[2(at\, \textrm{mod}\, 1)-1],
\label{eq:1.8}
\end{equation}
panels (d)--(e) represent the time-series for a uniform random
distribution of the delay times $T(t)$ in an interval
$[T_0-\varepsilon,T_0+\varepsilon]$:
\begin{equation}
T(t)=T_0+\varepsilon\,\textrm{Random}\,[-1,1],
\label{eq:1.9}
\end{equation}
and panels (f)--(g) are related to periodic sine-wave modulation:
\begin{equation}
T(t)=T_0+\varepsilon\,\sin{(at)}.
\label{eq:1.10}
\end{equation}
In each case, the control parameters were chosen as $K=0.5$,
$T_0=30$ and $\varepsilon=1$, with $a=10$ for the sawtooth and 
sine waves. We note that the values for the control parameter 
$\varepsilon$ are limited to the interval $0\leq\varepsilon\leq T_0$. 
In panels (c), (e) and (g) we see that the feedback signal $F(t)$ 
vanishes when the stabilization of the fixed point is achieved, suggesting
noninvasiveness of VDFC. The noninvasiveness of the control method
follows from the form of the control force in Eq. (\ref{eq:1.4}),
since $y(t-T(t))=y(t)$ if the fixed point is stabilized. We
notice that the saddle point $C_0$ at the origin cannot be
stabilized with the VDFC scheme, obeying the odd-number limitation
theorem that a Pyragas-type control in its standard form is
limited to fixed points characterized by a finite torsion, which
do not have an odd number of positive real eigenvalues
\cite{JBO97,NAK97,NU98,FFG07}. The fixed points $C^\pm$ do not
have positive real eigenvalues, and they can be stabilized with
the proposed version of VDFC, as indicated in Fig. 2. Since the
fixed points $C^+$ and $C^-$ have identical eigenvalues, the
Lorenz system can be stabilized to either of these states,
depending on the initial conditions. In the case of uniform random
modulation (\ref{eq:1.9}), the preference of control toward either
$C^+$ or $C^-$ will also depend on the random number sequence used
in the modulation. In the latter case, we noticed that for
identical initial conditions and different uniform random
sequences, the preference of stabilization of the system varies
between $C^+$ and $C^-$. Also, keeping the random sequence fixed
and varying the starting position of the sequence influences the
preference of the control. These observations could be related to
the structure of the basins of attraction for the stabilized
steady states $C^\pm$.

The simulation was repeated for the R\"{o}ssler system \cite{ROS76}:
\begin{eqnarray}
{d\over dt}x(t)&=&-y(t)-z(t), \label{eq:ros1}\\
{d\over dt}y(t)&=&x(t)+0.2\, y(t)+F(t), \label{eq:ros2}\\
{d\over dt}z(t)&=&0.2\, +(x(t)-5.7)z(t), \label{eq:ros3}
\end{eqnarray}
with the feedback force $F(t)$ as
given in Eq. (\ref{eq:1.4}), using the same types of time
modulation of the delay time $T(t)$. The unperturbed R\"{o}ssler
system has two unstable fixed points of the focus type at 
$C_1(0.007,-0.035,0.035)$ and $C_2(5.693,-28.465,28.465)$ 
with corresponding eigenvalues 
$\lambda(C_1)\approx\{-5.687,0.097+0.995i,0.097-0.995i\}$ and
$\lambda(C_2)\approx\{0.192,-4.596\cdot10^{-6}+5.428i,-4.596\cdot10^{-6}-5.428i\}$.
The stabilization has been achieved only for the fixed point
$C_1$, since it doesn't possess positive real eigenvalues, whereas
the fixed point $C_2$ has one, and therefore, cannot be stabilized
with the present version of VDFC. We omit a detailed discussion on
the stabilization process and the simulation figures describing
the dynamics of the control, since they concur to the ones for the
Lorenz system.

\section{The mechanism of VDFC}

To provide an insight into the mechanism of the VDFC method
and to reveal its superiority over the classical TDAS, 
we will initially consider a two-dimensional
nonlinear system of first-order differential equations:
\begin{equation}
{d\over dt}\mathbf{x}(t)=\mathbf{f}[\mathbf{x}(t)],
\label{eq:2.1}
\end{equation}
with $\mathbf{x}(t)=\textrm{Col}[x(t),y(t)]$ being the state
column vector of the system, and $\mathbf{f}$ the field vector
describing the dynamics of the system. The stability of a
particular critical point $\mathbf{\overline{x}}$ can be
determined by linearizing the vector field $\mathbf{f}$ around
$\mathbf{\overline{x}}$. We assume that the system (\ref{eq:2.1})
has undergone a coordinate transformation, such that the critical
point $\mathbf{\overline{x}}$ is at the origin
($\mathbf{\overline{x}=0}$). We also assume that the critical
point $\mathbf{\overline{x}=0}$ is an unstable fixed point of
focus type. Here we have taken into account the limitation of the
original Pyragas scheme, suggesting a failure of the control
method for a torsion-free situation, i. e. if the number of
positive real eigenvalues of the unstable fixed point is odd. The
linearized version of the system (\ref{eq:2.1}) in center manifold
coordinates can be written as:
\begin{equation}
{d\over dt}\mathbf{x}(t)=\mathbf{A}\mathbf{x}(t),
\label{eq:2.2}
\end{equation}
where
\begin{equation}
\mathbf{A}=
\left(
\begin{array}{cc}
\lambda & \omega \\
-\omega & \lambda \\
\end{array}
\right)
\label{eq:2.3}
\end{equation}
is the matrix that determines the dynamics of the unperturbed
system. The matrix $\mathbf{A}$ has complex conjugate eigenvalues
$\Lambda_0=\lambda\pm i\,\omega$, with $\lambda$ and $\omega$
positive real numbers, warranting an unstable focus at the origin.

To stabilize the unstable fixed point at the origin using VDFC, we
will perturb the system (\ref{eq:2.1}) with an additional control
force $\mathbf{F}(t)$ in the diagonal form:
\begin{equation}
\mathbf{F}(t)=
K\left(
\begin{array}{c}
x(t-T(t))-x(t) \\
y(t-T(t))-y(t)
\end{array}
\right),
\label{eq:2.4}
\end{equation}
where $K$ is the feedback strength, and $T(t)$ is the
time-dependent delay time. The linearized system (\ref{eq:2.2})
now obtains the form:
\begin{equation}
{d\over dt}\mathbf{x}(t)=\mathbf{A}\mathbf{x}(t)+\mathbf{F}(t),
\label{eq:2.5}
\end{equation}
which, in the case when $T(t)$ is constant, reduces itself to the
standard diagonal TDAS control scheme. In the following, we
consider a modulated time-delay $T(t)$ around a nominal value
$T_0$, in a form of a periodic sawtooth wave (\ref{eq:1.8}). In
this case, the delay time $T(t)$ is uniformly distributed over the
interval $[T_0-\varepsilon,T_0+\varepsilon]$. According to
Michiels-Van Assche-Niculescu (MAN) theorem \cite{MAN05}, the
stability of ($\ref{eq:2.5}$) under the variable-delay control
force (\ref{eq:2.4}) can be inferred from the stability of the
analogous time-invariant system with a distributed delay:
\begin{equation}
{d\over dt}\mathbf{x}(t)=\mathbf{A}\mathbf{x}(t)+\mathbf{\widetilde{F}}(t),
\label{eq:2.6}
\end{equation}
having a control force $\mathbf{\widetilde{F}}(t)$ in the form:
\begin{equation}
\mathbf{\widetilde{F}}(t)=
K\left(
\begin{array}{c}
\displaystyle{{1\over 2\varepsilon}\int_{t-T_0-\varepsilon}^{t-T_0+\varepsilon} x(\theta)\,d\theta-x(t)}
\vspace{0.5cm}\\
\displaystyle{{1\over 2\varepsilon}\int_{t-T_0-\varepsilon}^{t-T_0+\varepsilon} y(\theta)\,d\theta-y(t)}
\end{array}
\right).
\label{eq:2.7}
\end{equation}
The theorem asserts that if the comparison system
(\ref{eq:2.6})--(\ref{eq:2.7}) is asymptotically stable, then the
original system (\ref{eq:2.4})--(\ref{eq:2.5}) is globally
uniformly asymptotically stable for sufficiently large values of
the inverse period $a$ of the modulation. We note that in the case
of a non-uniform deterministic distribution of $T(t)$ in the
interval $[T_0-\varepsilon,T_0+\varepsilon]$, an additional
multiplicative factor will appear in the integrals in Eq.
($\ref{eq:2.7}$) depending on the form of $T(t)$, due to the
non-constant weight of the distribution.

Using the ansatz $\mathbf{x}(t)\sim \exp(\overline{\Lambda} t)$ in
Eq. (\ref{eq:2.6}), we obtain the characteristic equation for the
eigenvalues $\overline{\Lambda}$ of the comparison system
(\ref{eq:2.6})--(\ref{eq:2.7}):
\begin{equation}
\lambda\pm i\omega=\overline{\Lambda}+K\, \left(1-{\sinh{(\overline{\Lambda}\,\varepsilon)}\over
\overline{\Lambda}\,\varepsilon}\,e^{-\overline{\Lambda} T_0}\right).
\label{eq:2.11}
\end{equation}
Under the conditions of the MAN theorem, the stability of the
original system (\ref{eq:2.4})--(\ref{eq:2.5}) is determined by
the roots $\overline{\Lambda}$ of the characteristic equation
(\ref{eq:2.11}), providing that the frequency $a$ of variation of
the delay is large compared to the system's dynamics. In this
sense, the roots $\overline{\Lambda}$ may be considered as
effective eigenvalues describing the overall stability of the
original variable-delay system.

The presence of the parameters $\varepsilon$ and $T_0$ determining
the delay interval makes the characteristic equation
(\ref{eq:2.11}) transcendental in $\overline{\Lambda}$, possessing
countable infinite set of complex solutions. The control is
successful if for some $\varepsilon$, $K$ and $T_0$ the real parts
of all the eigenvalues $\overline{\Lambda}$ are negative.

An alternative heuristic way to arrive to Eq. (\ref{eq:2.11}) is
the following. We shall look for a solution of Eq.
($\ref{eq:2.5}$) in the form $\mathbf{x}(t)\sim \exp(\Lambda(t)
t)$. The equation for $\Lambda(t)$ is
\begin{equation}
\dot{\Lambda}t+\Lambda=\lambda\pm i\omega+K
\left(e^{\left[\Lambda(t-T(t))-\Lambda(t)\right)]t-\Lambda(t-T(t))T(t)}-1\right).
\label{RavenkaLambda}
\end{equation}
Considering the asymptotic domain $t\gg1$ (and therefore $t\gg
T(t)$), the finite variation of the delay-time $T(t)$ and the
approximation $\Lambda(t-T(t))-\Lambda(t)\approx-\dot\Lambda(t)
T(t)$, from Eq. (\ref{RavenkaLambda}), one can conclude that there is an asymptotic
relationship $\dot\Lambda(t)\sim1/t$, implying that asymptotically
$\Lambda(t)$ becomes a constant. Integrating (\ref{RavenkaLambda}) over a period of
the time delay function $T(t)$ in the asymptotic domain of large
$t$ results in Eq. (\ref{eq:2.11}).

We numerically analyze Eq. (\ref{eq:2.11}) to obtain the domains
of stability of the fixed point in the parameter plane $K-T_0$ for
different values of $\varepsilon$, keeping the parameters of the
unstable focus fixed at $\lambda=0.5$ and $\omega=\pi$. The
results are shown in panels (a) through (d) of Fig. 3,
corresponding to $\varepsilon=0$, 0.3, 0.5 and 1 respectively.

\begin{figure*}
\includegraphics[width=\textwidth,height=!]{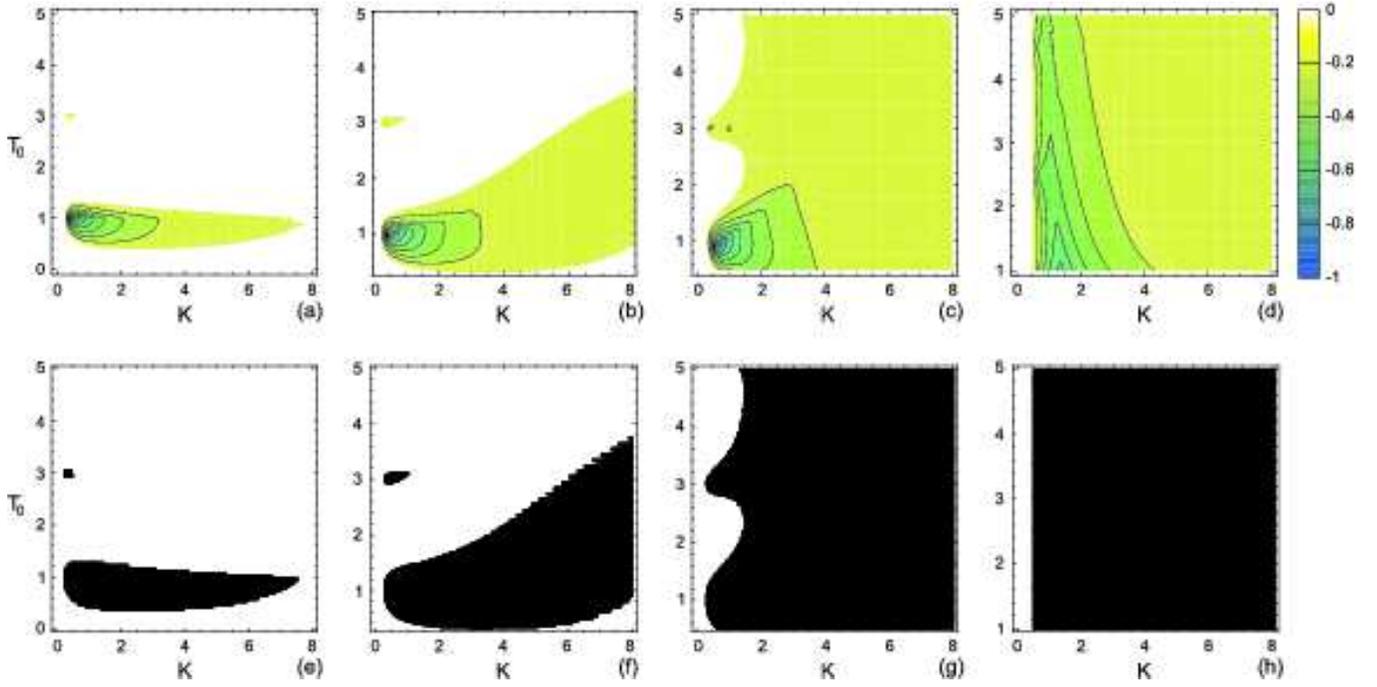}
\caption{(Color online) (a)--(d) Domains of control in the
$(K,T_0)$ plane for a sawtooth-wave modulation of the delay time,
obtained by a numerical solution of the characteristic equation
(\ref{eq:2.11}). The values of the modulation amplitude are: (a)
$\varepsilon=0$, (b) $\varepsilon=0.3$, (c) $\varepsilon=0.5$, (d)
$\varepsilon=1$. For combinations of $K$ and $T_0$ belonging to
the shaded areas, the largest real part of the complex eigenvalues
$\overline{\Lambda}$ is negative, and the control is successful.
The parameters of the unstable focus are $\lambda=0.5$ and
$\omega=\pi$. (e)--(h) The corresponding stability regions (black
areas) determined by a computer simulation of VDFC for the system
(\ref{eq:2.5}). Note the shifts of the origin along the $T_0$-axis
by an amount equal to $\varepsilon$.}
\end{figure*}

The shaded areas correspond to the set of control parameters
$(K,T_0)$ for which the largest real part of the complex
eigenvalues $\overline{\Lambda}$ is negative, indicating a
successful control. For combinations $(K,T_0)$ belonging to the
white area, $\textrm{max}[\textrm{Re}(\overline{\Lambda})]>0$, and
the control is not possible. The values of
$\textrm{max}[\textrm{Re}(\overline{\Lambda})]$ are given by the
grayscale in the upper right corner of Fig. 3. The control
becomes more robust for larger magnitude of the negative
$\textrm{max}[\textrm{Re}(\overline{\Lambda})]$. When
$\varepsilon=0$, the VDFC method reduces to the standard TDAS
control method, and the domain of successful control is indicated
in panel (a) of Fig. 3 \cite{HS05}.  As $\varepsilon$ becomes
larger than zero and closer to 1 (panels (b) through (d)), the
domain of control is drastically enlarged, reaching its maximum at
$\varepsilon=1$ (panel (d)). We note that due to the limitation of
the control parameter $\varepsilon$ to the interval
$0\leq\varepsilon\leq T_0$, the origin of the axis $T_0$ is taken
at $T_0=\varepsilon$.

In panels (e) through (h) of Fig. (3) we show the stability
regions corresponding to panels (a) through (d), obtained by a
computer simulation of the VDFC control scheme (\ref{eq:2.5}). The
simulation is performed by picking a point from the parameter
plane $K-T_0$ and numerically integrating the system
(\ref{eq:2.5}). The time-modulation of $T(t)$ used in the
simulation was in a form of a sawtooth wave ($\ref{eq:1.8}$) with
an inverse period $a=10$. The combinations $(K,T_0)$ that lead to
a successful fixed point stabilization are marked in black. We
immediately notice an excellent resemblance between the stability
domains obtained by computer simulation and by numerically
analyzing the characteristic equation (\ref{eq:2.11}). The
resemblance may be further improved by requiring a larger accuracy
of the Runge-Kutta method used in the simulation.

To obtain an analytical description of the boundaries of the
control domain in Fig. 3, we substitute
$\overline{\Lambda}=\Pi+i\,\Omega$ into the characteristic
equation (\ref{eq:2.11}) and separate the equation into real and
imaginary parts. Since $\Pi=0$ at the threshold of control, the
resulting equations are reduced to:
\begin{eqnarray}
\lambda&=&K\left(1-{\sin{(\Omega\varepsilon)}\over \Omega\varepsilon}\cos{(\Omega T_0)}\right),
\label{eq:2.12}\\
\pm\omega&=&\Omega+K{\sin{(\Omega\varepsilon)}\over \Omega\varepsilon}\sin{(\Omega T_0)},
\label{eq:2.13}
\end{eqnarray}
which can be algebraically manipulated to obtain a parametric
representation of the dependence on $\Omega$ of the boundary of
the successful control domain
\begin{widetext}
\begin{eqnarray}
K(\Omega)&=&{\lambda\pm\sqrt{\lambda^2-\{1-[\sin{(\Omega\varepsilon)}/(\Omega\varepsilon)]^2\} [\lambda^2+(\omega-\Omega)^2]} \over 1-[\sin{(\Omega\varepsilon)}/(\Omega\varepsilon)]^2},\label{eq:long1}\\
T_0(\Omega)_1&=&{1\over\Omega}\left[2n\pi+\arccos\left(
\lambda[\sin{(\Omega\varepsilon)}/(\Omega\varepsilon)]^2\pm\sqrt{\lambda^2-\{1-[\sin{(\Omega\varepsilon)}/(\Omega\varepsilon)]^2\} [\lambda^2+(\omega-\Omega)^2]} \over
\left[\sin{(\Omega\varepsilon)}/(\Omega\varepsilon)\right]
\left[\lambda\pm\sqrt{\lambda^2-\{1-[\sin{(\Omega\varepsilon)}/(\Omega\varepsilon)]^2\} [\lambda^2+(\omega-\Omega)^2]}\right]
\right)\right],\label{eq:long2}\\
T_0(\Omega)_2&=&{1\over\Omega}\left[2(n+1)\pi-\arccos\left(
\lambda[\sin{(\Omega\varepsilon)}/(\Omega\varepsilon)]^2\pm\sqrt{\lambda^2-\{1-[\sin{(\Omega\varepsilon)}/(\Omega\varepsilon)]^2\} [\lambda^2+(\omega-\Omega)^2]} \over
\left[\sin{(\Omega\varepsilon)}/(\Omega\varepsilon)\right]
\left[\lambda\pm\sqrt{\lambda^2-\{1-[\sin{(\Omega\varepsilon)}/(\Omega\varepsilon)]^2\} [\lambda^2+(\omega-\Omega)^2]}\right]
\right)\right],
\label{eq:long3}
\end{eqnarray}
\end{widetext}
where $n$ is a nonnegative integer characterizing different leaves
arising from the multivaluedness of the arccosine function.

When $\varepsilon=0$ [see panels (a) and (e) in Fig. 3], VDFC is
reduced to TDAS, and Eqs. (\ref{eq:2.12}) and (\ref{eq:2.13}) read
\cite{HS05}:
\begin{eqnarray}
\lambda&=&K\left(1-\cos{(\Omega T_0)}\right),
\label{eq:2.17}\\
\pm\omega&=&\Omega+K\sin{(\Omega T_0)}.
\label{eq:2.18}
\end{eqnarray}
In this case, when the product $\Omega T_0$ is an odd multiple of
$\pi$, we obtain $\Omega=\omega$ and $K=K_{min}=\lambda/2$. Hence,
the points $K=K_{min}=\lambda/2$ and $T_0=(2n+1)\pi/\omega$
correspond to points of successful control in the $K-T_0$ plane
with minimal feedback gain. Similarly, when $\Omega T_0$ is an
even multiple of $\pi$, that is, when $T_0=2n\pi/\omega$, the
control fails for any feedback gain. The domain of control
consists of stability islands at $T_0$ corresponding to odd $n$
isolated by regions encompassing $T_0$ corresponding to even $n$
for which the control fails.

When $\varepsilon>0$ [panels (b)--(d) and (f)--(h) in Fig. 3],
from Eqs. (\ref{eq:2.12}) and (\ref{eq:2.13}) we obtain:
\begin{equation}
K=K_{min}={\lambda\over 1+\sin{(\omega\varepsilon)}/(\omega\varepsilon)}
\end{equation}
for $\Omega T_0=(2n+1)\pi$, and:
\begin{equation}
K=K_{min}={\lambda\over 1-\sin{(\omega\varepsilon)}/(\omega\varepsilon)}
\end{equation}
for $\Omega T_0=2n\pi$, implying considerable reconfiguration of
the stability islands as soon as $\varepsilon>0$, which is readily
observable from Fig. 3. Specifically, if $\varepsilon=1$ and
$\omega=\pi$, then $K_{min}=\lambda$ for any integer $T_0$, as is
clearly indicated from panels (d) and (h) in Fig. 3.

\begin{figure*}
\includegraphics[width=\textwidth,height=!]{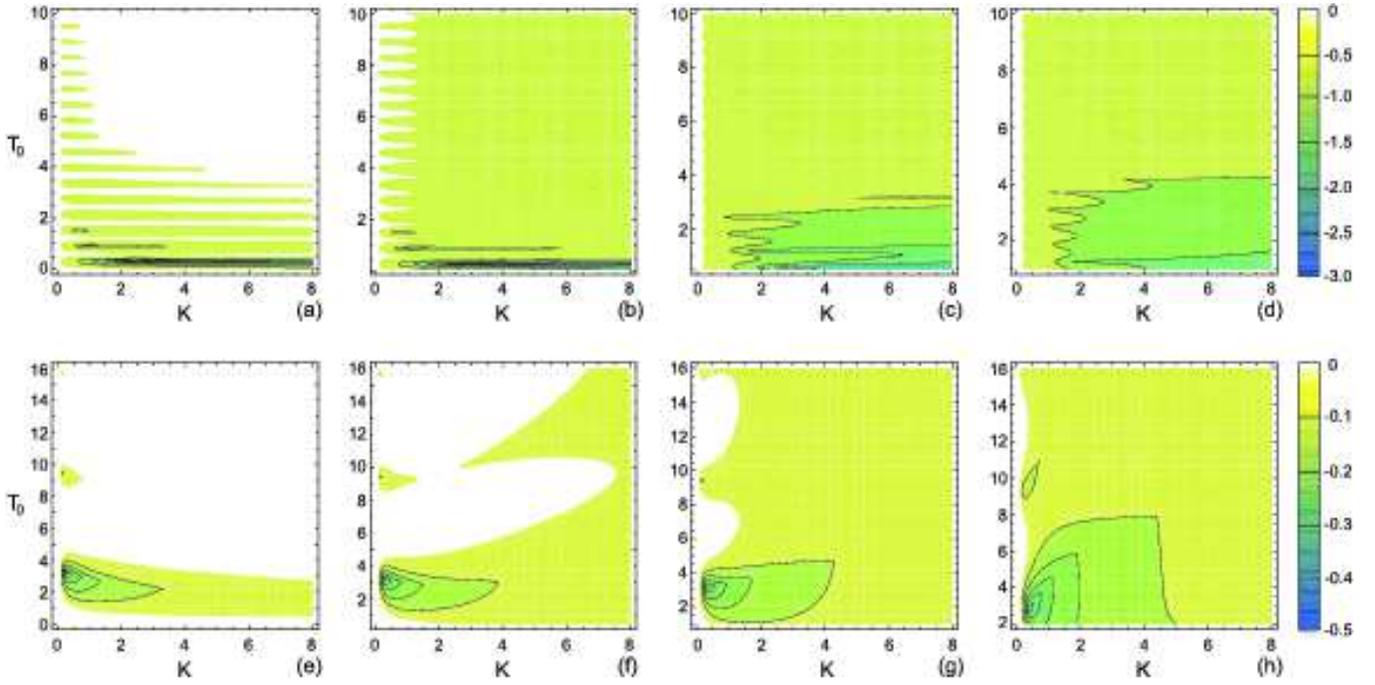}
\caption{(Color online) (a)--(d) Domains of successful control in 
the $(K,T_0)$ plane for the unstable steady states $C^{\pm}$ 
in the Lorenz system (\ref{eq:lor1})--(\ref{eq:lor3}). 
The modulation of the delay time is in a form of a sawtooth-wave 
(\ref{eq:1.8}), and the values of the modulation amplitude are: 
(a) $\varepsilon=0$, (b) $\varepsilon=0.1$, 
(c) $\varepsilon=0.5$, (d) $\varepsilon=1$. 
(e)--(h) The corresponding domains of control for the unstable 
fixed point $C_1$ in the R\"{o}ssler 
system (\ref{eq:ros1})--(\ref{eq:ros3}). 
In this case, the values of the modulation amplitude are: 
(e) $\varepsilon=0$, (f) $\varepsilon=0.5$, (g) $\varepsilon=1$, 
(h) $\varepsilon=2$. We note that the numerical calculations 
for the remaining two unstable states in the Lorenz and R\"{o}ssler 
systems show no regions of successful control, in accordance with our
previous discussion related to the odd-number limitation theorem.}
\end{figure*}

To conclude this section, we numerically calculated the parameter regions of 
the effective fixed point control for the Lorenz system (\ref{eq:lor1})--(\ref{eq:lor3}) 
and the R\"{o}ssler system (\ref{eq:ros1})--(\ref{eq:ros3})
in the case of a periodic sawtooth-wave modulation (\ref{eq:1.8}). 
The calculations were performed for different values of the 
modulation amplitude $\varepsilon$, following the same strategy 
as for the two-dimensional case discussed previously. We omit 
showing the resulting characteristic equations and the parametric 
formulas describing the boundary of the domain of control due to 
their length and complexity. The obtained stability regions are 
depicted in Fig. (4). 

\section{Conclusion}

We have demonstrated that implementation of a time-varying delay
into the standard TDAS control scheme can dramatically improve the
efficiency of control of unstable steady states. The method of
variable delay feedback control (VDFC) was illustrated for the
paradigmatic Lorenz and R\"{o}ssler systems, showing the usual
control failure of TDAS in the case of unstable steady states
having an odd-number of positive real eigenvalues. We expect that
this limitation may be prevailed by a suitable choice of the
feedback control matrix $K$ \cite{FFG07,JFG07,PS07}, or by using
the VDFC method in combination with more sophisticated control
schemes \cite{SS97,P01}.

Using Michiels-Van Assche-Niculescu theorem \cite{MAN05} and an
independent approach for the case of a time-varying delay in a
form of a sawtooth-wave, we were able to perform a linear
stability analysis to calculate the domain of control for a
two-dimensional unstable focus in the plane parametrized by the
feedback gain $K$ and the nominal time-delay $T_0$ for different
values of the parameter $\varepsilon$ determining the amplitude of
the modulation. The same analysis was repeated for the three-dimensional
Lorenz and R\"{o}ssler systems. In parallel to the analytic derivation, we
performed computer simulations that confirmed the results of the
analytical approach. In this way, we showed that variable delay
feedback control allows stabilization of unstable steady states
over much larger domain of parameters in comparison to the usual TDAS
control scheme.

Various modifications and extensions of the VDFC method, like using
different forms of delay modulations, including a variable time-delay into
the ETDAS scheme and other schemes, the influence of the nonzero control-loop
latency \cite{HS05}, bandpass filtering and non-diagonal coupling, as well 
as experimental implementation of the VDFC, are all subjects of
ongoing analysis.

\end{document}